\documentclass[a4paper,11pt]{article}

\usepackage{amsmath}
\usepackage{amssymb}
\usepackage{color}
\usepackage[dvips]{graphicx}
\usepackage{cite}
\usepackage{hyperref}

\makeatletter
\@addtoreset{equation}{section}
\renewcommand{\theequation}{\thesection.\@arabic\c@equation}
\makeatother

\makeatletter
\renewcommand\appendix{\par%\newpage
  \setcounter{section}{0}%
  \setcounter{subsection}{0}%
  \gdef\thesection{Appendix \@Alph\c@section }
  \renewcommand{\theequation}
  {\Alph{section}.\arabic{equation}}
}
\makeatother

\def \be {\begin{equation}}
\def \ee {\end{equation}}
\def \ba {\begin{array}}
\def \ea {\end{array}}
\def \bea{\begin{eqnarray}}
\def \eea{\end{eqnarray}}

\def \l {\lambda}
\def \L {\Lambda}

\def \r {\rho}
\def \o {\omega}

\def \f {\frac}

\def \la {\leftarrow}

\def \la {\label}
\def \fn {\footnote}

\title{\textbf{Dark energy and normalization of cosmological wave function in modified gravitations}}

\author{
Peng Huang$^{1}$ \fn{hp@zcmu.edu.cn} ,\,
Yue Huang $^{2,3}$ \fn{huangyue@itp.ac.cn}
}

\date{}

\begin{document}

\maketitle

\begin{center}
{\it
$^{1}$ Department of Information, Zhejiang Chinese Medical University, \\
 Hangzhou 310053, China \\
\vspace{2mm}
$^{2}$ Key Laboratory of Theoretical Physics, Institute of Theoretical Physics,\\
Chinese Academy of Sciences, \\
Beijing 100190, China \\
\vspace{2mm}
$^{3}$ School of Physical Sciences, University of Chinese Academy of Sciences,
 \\
No.19A Yuquan Road, Beijing 100049, China}
\vspace{10mm}
\end{center}

\begin{abstract}
Based on Wheeler-DeWitt equation derived from general relativity, it had been found that only dark energy can lead to a normalizable cosmological wave function.
It is shown in the present work that, for dRGT gravity, Eddington-inspired-Born-Infeld gravity and Ho$\check{\text{r}}$ava-Lifshitz gravity, the previous conclusion can also stand well in quantum cosmology induced from these modified gravities. This observation implies that there might be a universal relation between dark energy and normalizability of the cosmological wave function.

\end{abstract}

\baselineskip 18pt

\thispagestyle{empty}

\newpage

\section{Introduction}\label{section1}

The mysterious accelerating expansion of the current universe is of central concern both in theoretical and experimental aspects since its discovery \cite{Perlmutter:1998np,Riess:1998cb}. The unknown reason that triggers the accelerating expansion is dubbed as dark energy. Despite of nearly two decades of intense study, it is fair to say that the intrinsic nature of the dark energy is not well understood. A satisfied solution to dark energy problem may need a complete knowledge of quantum gravity which is still under investigation. In this situation, trying new perspective into dark energy problem is necessary.

When expected to be an exotic energy component, dark energy can be well described with the help of the holographic principle \cite{Li:2004rb}, a characteristic feature of any viable theory of quantum gravity. On the other hand, as an application of quantum physics to the classical cosmology, quantum cosmology is also tightly related to (or originates from) quantum gravity. Since both dark energy and quantum cosmology have connections to quantum gravity, it is is well and naturally motivated to investigate dark energy from the perspective of quantum cosmology. A trial along this line of thought was done in \cite{Huang:2015eua}. It was found that only dark energy (an exotic energy component) can give rise to a normalizable cosmological wave function. This result indicates that, while being responsible for the classical accelerating expansion of the universe, dark energy can also lead to very different behaviors at the quantum level compared with ordinary energy components.
%The significant difference between dark energy and ordinary energy components from the perspective of quantum cosmology may imply interesting physics and needs further investigation.

What needs to be emphasized is that the previous conclusion---only dark energy can lead to a normalizable cosmological wave function---is obtained based on the hypothesis that dark energy is an exotic energy components. However, it is completely possible that dark energy comes from a modification of general relativity on cosmological scale. The key point is, while energy component show itself in Wheeler-DeWitt (WD) equation as a potential term (function of scale factor $a$ multiply by cosmological wave function $\Psi(a)$), modified gravitation may change the dynamical part (differential of the cosmological wave function with respect to $a$ ) of WD equation besides introducing nontrivial exotic potential term. Therefore, even though that a suitable exotic energy or an appropriate modification of general relativity can achieve the same purpose on explaining the accelerating expansion of the current universe, the WD equations attached to this two different approach may differ from each other distinctly.

Then comes the question: Does the conclusion that only dark energy (treated as an exotic energy component) can lead to a normalizable cosmological wave function also stand well when dark energy is an effect of modified gravitation? At first glance, the answer is more likely to be negative since WD equations for these two different cases, as explained previously, could be very different to each other. However, it is shown in present work that, for some representative modified gravitations, dark energy also leads to a normalizable cosmological wave function, which can be regarded as strong clue for the observation that dark energy leads to a normalizable cosmological wave function is a universal conclusion. This is the main content of the present work.

The paper is organized as following: Sec.(\ref{section2}) is devoted to a brief review of work that had been done in \cite{Huang:2015eua}  for later  convenience. In Sec.(\ref{section3}), three representative modified theories of gravity are investigated in detail. One can see that, while WD equation induced from dRGT gravity has a very different dynamical part comparing to the ordinary one (derived from GR), new potential terms will be introduced in WD equations derived from Eddington-inspired-Born-Infeld gravity and Ho$\check{\text{r}}$ava-Lifshitz (HL) gravity. However, in all the three representative modified theories of gravity, the conclusion that only dark energy can lead to a normalizable cosmological wave function stands well.

\section{Dark energy in general relativity: a quantum cosmological perspective}\la{section2}

Dark energy is investigated from the perspective of quantum cosmology derived from general relativity in \cite{Huang:2015eua}. For a universe filled only with the cosmological constant, the Wheeler-DeWitt equation is
\begin{equation}
\label{3}
[-\frac{\partial^{2}}{\partial a^{2}}-\frac{q}{a}\frac{\partial
}{\partial a}+\frac{1}{l_p^4}( k a^{2}-\frac{a^{4}}{l_{\Lambda}^2}
)]\Psi(a)=0
\end{equation}
with $k$, $l_p=(\frac{4G^{2} \hbar^2}{9\pi^{2}})^{\frac{1}{4}}$, $l_{\Lambda}=(\frac{3}{\Lambda})^{\frac{1}{2}}$ corresponding to  the spatial curvature,  Planck scale and length scale introduced by the cosmological constant. In this WD equation, energy components coming from spatial curvature and cosmological constant are reflected in term  $\frac{k a^{2}}{l_p^4}$ and $\frac{a^{4}}{l_p^4l_{\Lambda}^2}$ respectively.
Then, similar to quantum mechanics, the inner product of the cosmological wave function is defined as
\begin{equation}
\label{2}
\mathcal{P}=\int_{0}^{+\infty} {a^q|\Psi|^2da}.
\end{equation}
The integral $\mathcal{P}$ can be divided as
\begin{equation}
\label{567}
\mathcal{P}=\int_{0}^{\epsilon} {a^q|\Psi|^2da}+\int_{\epsilon}^{C} {a^q|\Psi|^2da}+\int_{C}^{+\infty} {a^q|\Psi|^2da},
\end{equation}
the potentially singular behavior of the cosmological wave function comes from the first and the third part. When $\mathcal{P}$ is finite, $\Psi$ is said to be normalizable otherwise non-normalizable. Solving differential equation (\ref{3}) and inserting its solution into (\ref{567}), it can be found that normalizability of the cosmological wave function imposes strict constraints on the form of WD equation: on one hand, the normal ordering ambiguity factor $q$ must take value in the domain of (-1, 3) to ensure no divergence in $\int_{0}^{\epsilon} {a^q|\Psi|^2da}$; on the other hand, the existence of cosmological constant (the appearance of term $\frac{a^{4}}{l_p^4l_{\Lambda}^2}$) is obligatory to ensure the convergence of $\int_{C}^{+\infty} {a^q|\Psi|^2da}$.

To see whether the present result has a more universal meaning, the WD equation (\ref{3}) is generalized to the following form
\begin{equation}
\label{6}
[-\frac{\partial^{2}}{\partial a^{2}}-\frac{q}{a}\frac{\partial
}{\partial a}+\frac{1}{l_p^4}( a^{2}-\frac{1}{f(h)} \frac{a^h}{l_p^{h-2}})]\Psi(a)=0,
\end{equation}
in which $f(h)$ is a general function of parameter $h$ and $f(4)=(\frac{l_{\Lambda}}{l_p})^2$ is required. By exploring the classical limit of Eq.(\ref{6}), one can see that term $\frac{1}{f(h)} \frac{a^h}{l_p^{h-2}}$ in fact corresponds to a general energy component with index of equation of state having the value of $w=\frac{1-h}{3}$. Apparently, $h=0, 1, 2, 4$ corresponds to radiation, matter, energy coming from spatial curvature and cosmological constant, respectively. Then it is shown that, for small $a$, the divergence in $\int_{0}^{\epsilon} {a^q|\Psi|^2da}$ can always be avoided by choosing an appropriate normal ordering ambiguity factor $q$ in region $(-1,3)$; for large $a$, only when $h>2$ (i.e., $w<-\frac{1}{3}$) can there be no divergence in $\int_{C}^{+\infty} {a^q|\Psi|^2da}$.

Through the above analysis, one has the following conclusion: equipped with an appropriate normal ordering ambiguity factor $q$, only dark energy can lead to a normalizable cosmological wave function. One should notice that such conclusion is obtained on the assumption that dark energy is not an effect of modified gravitation but an exotic energy component.

\section{Dark energy in modified gravitations}\la{section3}
\subsection{dRGT gravity}

dRGT is a healthy massive gravity in which no ghosts exist \cite{deRham:2010ik,deRham:2010kj}.  Suppose the metric has the following form
\begin{equation}
g_{\mu\nu}=\eta_{\mu\nu}+h_{\mu\nu}=\eta_{ab}\partial_{\mu}\phi^a(x)\partial_{\nu}\phi^b(x)+H_{\mu\nu},
\end{equation}
then the action of dRGT is
\begin{equation}\label{equ:Action_f_dRGT}
S_g=\int d^4x\sqrt{-g}[R-\frac{m^2}{4}\mathcal{U}(g,H)]
\end{equation}
in which the first term comes from Einstein-Hilbert action and the second term manifests the modification caused by a small but nonzero mass of graviton. $\mathcal{U}(g,H)$ can be expressed as
\begin{equation}
\mathcal{U}(g,H)=-4(\mathcal{L}_2+\alpha_3\mathcal{L}_3+\alpha_4\mathcal{L}_4)
\end{equation}
with $\mathcal{L}_2$, $\mathcal{L}_3$ and $\mathcal{L}_4$ defined as following:
\begin{equation}
\left\lbrace
\begin{aligned}
&\mathcal{L}_2=\frac{1}{2}(<\mathcal{K}>^2-<\mathcal{K}^2>),\\
&\mathcal{L}_3=\frac{1}{6}(<\mathcal{K}>^3-3<\mathcal{K}><\mathcal{K}^2>+2<\mathcal{K}^3>),\\
&\mathcal{L}_4=\frac{1}{24}(<\mathcal{K}>^4-6<\mathcal{K}>^2<\mathcal{K}^2>+3<\mathcal{K}^2>^2+8<\mathcal{K}><\mathcal{K}^3>-6<\mathcal{K}^4>),
\end{aligned}
\right.
\end{equation}
where $\mathcal{K}_{\mu\nu}$ stands for
\begin{equation}\label{equ:Def_o_K}
\mathcal{K}^{\mu}_{\nu}(g,H)\equiv\delta^{\mu}_{\nu}-\sqrt{\eta_{ab}\partial^{\mu}\phi^a\partial_{\nu}\phi^b}
\end{equation}
and symbol $<...>$ means trace, i.e., $<\mathcal{K}>=g^{\mu\nu}\mathcal{K}_{\mu\nu}$, $<\mathcal{K}^2>=g^{\rho\sigma}g^{\mu\nu}\mathcal{K}_{\rho\mu}\mathcal{K}_{\sigma\nu}$, ...\fn{For present purpose, it
suffices to content ourselves with the minimal introduction of dRGT, interested readers are referred to the original work in \cite{deRham:2010ik,deRham:2010kj}. The introduction of Eddington-inspired-Born-Infeld gravity and Ho$\check{\text{r}}$ava-Lifshitz gravity in later sections will also be in their simplest forms.}

Deriving WD equation from (\ref{equ:Action_f_dRGT}), a highly nontrivial task, had been carried out in \cite{Vakili:2012tm}:
\begin{equation}
\la{11}
\left[a^{-q}(-i\frac{\partial}{\partial a}+\frac{C_{\pm}m^2}{\sqrt{|k|}}a^3)[a^q(-i\frac{\partial}{\partial a}+\frac{C_{\pm}m^2}{\sqrt{|k|}}a^3)]-36|k|a^2-12c_{\pm}m^2a^4\right]\Psi(a)=0
\end{equation}
with $q$ the normal ordering ambiguity factor, $C_{\pm}$ and $c_{\pm}$  standing for
\bea
C_{\pm}&=& X_{\pm}(1-X_{\pm})[3+3\alpha_3(1-X_{\pm})+\alpha_4(1-X_{\pm})^2] \\
c_{\pm}&=&(X_{\pm}-1)[3(2-X_{\pm})+\alpha_3(1-X_{\pm})(4-X_{\pm})+\alpha_4(1-X_{\pm})^2]
\eea
with $X_{\pm} \equiv\frac{1+2\alpha_3+\alpha_4\pm\sqrt{1+\alpha_3+\alpha_3^2-\alpha_4}}{\alpha_3+\alpha_4}$.

To investigate the relation between energy components and normalization of cosmological wave function, introducing energy components into Eq.(\ref{11}) is necessary. Since the couple between matter and gravity is not impacted by massive graviton in dRGT, one can safely bring energy components into this theory as following:
\begin{equation}
\label{equ:WDW_o_MG}
\bigg[a^{-q}(-i\frac{\partial}{\partial a}+\frac{C_{\pm}m^2}{\sqrt{|k|}}a^3)[a^q(-i\frac{\partial}{\partial a}+\frac{C_{\pm}m^2}{\sqrt{|k|}}a^3)]-36|k|a^2-12c_{\pm}m^2a^4-\rho(a)a^4\bigg]\Psi(a)=0.
\end{equation}
This will be the start point of our discussion.

Comparing Eq.(\ref{equ:WDW_o_MG}) with Eq.(\ref{6}) in Sec.(\ref{section2}), it is apparent that the dynamical part of WD equation has changed dramatically, which may make solving the WD equation to be a difficult problem. Fortunately, one can circumvent this problem in dRGT: the central concern of the present work is whether the innerproduct (\ref{2}) is convergent, thus one has the freedom to rewrite the cosmological wave function $\Psi(a)$ as $\Psi(a)=e^{if(a)}\Phi(a)$. Let $f(a)=-i\frac{1}{4}\frac{C_{\pm}m^2}{\sqrt{|k|}}a^4$, WD equation (\ref{equ:WDW_o_MG}) turns into
\begin{equation}
\label{3.3}
\left[-a^{-q}(\frac{\partial}{\partial a}a^q\frac{\partial}{\partial a})-36|k|a^2-12c_{\pm}m^2a^4-\rho(a)a^4\right]\Phi(a)=0.
\end{equation}
It can be inferred from this WD equation that the square of the mass of graviton offers equivalently a nonzero cosmological constant to trigger the accelerating expansion of the universe, a characteristic feature of massive gravity. Similar to the discussions shown in Sec.(\ref{section2}), one can find that, for large $a$, the dominant term $-12c_{\pm}m^2a^4$ would lead to a normalizable cosmological wave function if there is no matter in the universe ($\r(a)=0$); what's more, in the limitation of $m=0$, $\r(a)$ (an energy component) must be dark energy otherwise the cosmological wave function cannot be normalizable. In other words, the conclusion that only dark energy can lead to a normalizable cosmological wave function also stands in dRGT.

\subsection{Eddington-inspired-Born-Infeld gravity}

Eddington-inspired-Born-Infeld gravity (EiBI) was proposed by Ba$\tilde{\text{n}}$ados and Ferreira as a combination of insights from Einstein and Eddington \cite{Banados:2010ix}. It is defined in Palatini formalism as
\begin{equation}\label{equ:Action_o_EiBI}
S_{EiBI}[g,\Gamma,\Phi]=\frac{2}{\kappa}\left[\sqrt{|g_{\mu\nu}+\kappa R_{\mu\nu}(\Gamma)|}-\lambda\sqrt{|g|}\right]+S_M[g,\Gamma,\Phi]
\end{equation}
with $\l$ a dimensionless parameter which cannot be zero because of self-consistency. Deriving WD equation from this action had been tactfully carried out in \cite{Bouhmadi-Lopez:2016dcf}. There the WD equation is
\begin{equation}\label{equ:WDW_f_EiBI}
\left[b^{-q}(\frac{\partial}{\partial b}b^q\frac{\partial}{\partial b})+\frac{48\lambda^3}{\kappa}b^4+\frac{24}{\kappa}(\lambda+\kappa\rho(a))^2a^6b^{-2}-\frac{72\lambda^2}{\kappa}a^2b^2\right]\Psi(b, a)=0.
\end{equation}
in which $\r=\r_0a^{-3(1+\o)}$ and $p$ are energy density and pressure, furthermore, $a$ and $b$ relate to each other through
\begin{equation}\label{equ:relation_o_a&b}
(\lambda+\kappa\rho)(\lambda-\kappa p)=\lambda^2 \frac{b^4}{a^4}.
\end{equation}
One can find that, new terms are introduced in the EiBI potential $\frac{48\lambda^3}{\kappa}b^4+\frac{24}{\kappa}(\lambda+\kappa\rho(a))^2a^6b^{-2}-\frac{72\lambda^2}{\kappa}a^2b^2$, which is more intricate than its counterpart in general relativity.

When $\r=p=0$, WD equation (\ref{equ:WDW_f_EiBI}) has a much more simpler form:
\begin{equation}
\left[b^{-q}(\frac{\partial}{\partial b}b^q\frac{\partial}{\partial b})+\frac{48\lambda^2(\lambda-1)}{\kappa}b^4\right]\Psi(b)=0.
\end{equation}
Comparing this equation with Eq.(\ref{3}), one can see that the intrinsic accommodation of dark energy in EiBI is reflected in the appearance of term $\frac{48\lambda^2(\lambda-1)}{\kappa}b^4$: $\l\neq 1$ corresponds to the existence of cosmological-constant-like dark energy, otherwise dark energy is eliminated in this theory. In case of $\l\neq 1$, as shown in Sec.(\ref{section2}), the cosmological wave function $\Psi(b)$ of EiBI is normalizable if the normal ordering ambiguity factor $q$ takes value in  domain of $(-1,3)$. In case of $\l=1$, the WD equation is
\begin{equation}
\left[b^{-q}(\frac{\partial}{\partial b}b^q\frac{\partial}{\partial b})\right]\Psi(b)=0.
\end{equation}
General solution of this equation is
\begin{equation}
\Psi(b)=\left\lbrace
\begin{aligned}
&C_1 b^{1-q}+C_2,\qquad q\neq1,\\
&C_1 \ln b+C_2, \qquad q=1.
\end{aligned}
\right.
\end{equation}
The corresponding integral $\mathcal{P}_2=\int_C^{\infty}b^q|\Psi|^2db$ ($C>0$) is
\begin{equation}
\int_C^{\infty}b^q|\Psi|^2db=\left\lbrace
\begin{aligned}
\int_C^{\infty}(C_1^2b^{2-q}+2C_1C_2b+C_2^2b^q) db,\qquad q\neq1,\\
\int_C^{\infty}b[C_1^2(\ln b)^2+2C_1C_2\ln b+C_2^2]db, \qquad q=1.
\end{aligned}
\right.
\end{equation}
Apparently these integrals cannot be convergent. The conclusion is that, assuming the absence of matter ($\r=p=0$), the cosmological wave function cannot be normalized if there is no dark energy.

When $\r \neq 0$, it is easy to obtain the following relations for large $a$ from (\ref{equ:relation_o_a&b}):
\begin{equation}
\left\lbrace
\begin{aligned}
&b^4\sim a^{-2-6\omega}, \quad\omega<-1\\
&b\sim a, \qquad\qquad \omega\geq-1.
\end{aligned}
\right.
\end{equation}
In case of $\o\geq-1$, $\r(b)=\r_0b^{-3(1+\o)}$ can be dark energy ($-\f{1}{3}>\o\geq-1$) or ordinary matter ($\o\geq-\f{1}{3}$), the WD equation corresponds to this situation is
\begin{equation}
\label{33}
\left[b^{-q}(\frac{\partial}{\partial b}b^q\frac{\partial}{\partial b})+\frac{48\lambda^2(\lambda-1)}{\kappa}b^4+48\lambda\rho(b)b^4+24\kappa\rho(b)^2b^4\right]\Psi(b)=0.
\end{equation}
If $\l\neq1$, the power of $b$ in dominant potential term in (\ref{33}) is definitely not smaller than $4$ which ensures the normalizability of the cosmological wave function. If $\l=1$, the build-in dark energy of EiBI is eliminated and the WD equation turns into
\begin{equation}
\left[b^{-q}(\frac{\partial}{\partial b}b^q\frac{\partial}{\partial b})+c\cdot b^{(1-3\omega)}\right]\Psi(b)=0.
\end{equation}
The normalization of wave function requires $1-3\omega>2$, i.e., $\omega<-1/3$. In case of $\o<-1$, the energy component $\r(a)$ must be dark energy. The WD equation is
\begin{equation}
\begin{aligned}
\left[b^{-q}(\frac{\partial}{\partial b}b^q\frac{\partial}{\partial b})+\frac{48\lambda^3}{\kappa}b^4+\frac{24\lambda^2}{\kappa}b^{-2-\frac{12}{1+3\omega}}+c_1\lambda b^{-\frac{8}{1+3\omega}}\right.\\
\left.+c_2\kappa b^{2-\frac{4}{1+3\omega}}-\frac{72\lambda^2}{\kappa}b^{2-\frac{4}{1+3\omega}}\right]\Psi(b)=0.
\end{aligned}
\end{equation}
The potential in this equation is complicated and it is difficult to eliminated the intrinsic dark energy introduced by EiBI. Thus there will always be dark energy in this situation.

The power of $b$ has four different values: $4$, $-2-\frac{12}{1+3\omega}$, $-\frac{8}{1+3\omega}$ and $2-\frac{4}{1+3\omega}$. They cannot equal to each other when when $\o<-1$. When $b$ is large enough, $b^4$ is the dominant term in these terms. According to discussions shown in Sec.(\ref{section2}), one can see that the inner product of the wave function is convergent. Thus, for modified gravity EiBI, it is also only dark energy can lead to a normalizable cosmological wave function.

\subsection{Ho$\check{\text{r}}$ava-Lifshitz gravity}

An important motivation for dRGT, as well as EiBI, is to explain the accelerating expansion of the universe from the perspective of modified gravitation. Nevertheless, the central focus of Ho$\check{\text{r}}$ava-Lifshitz (HL) gravity is to construct a theory of gravity that can be renormalizable \cite{Horava:2009uw},
no special attention has been paid to dark energy problem. It is interesting to see whether the conclusion---only dark energy can lead to a normalizable cosmological wave function---stands well in HL gravity, a theory whose primary goal is unrelated to dark energy. As a preliminary attempt, we will not consider the complicated coupling between matter and gravity in HL gravity \fn{See \cite{Wang:2017brl} for a clear overview on this subtle issue.}.

Starting with the projectable Ho$\check{\text{r}}$ava-Lifshitz gravity without detailed balance, its action is \cite{Sotiriou:2009gy,Sotiriou:2009bx}
\begin{equation}
\begin{aligned}
S_{HL}=\frac{M_{pl}^2}{2}\int d^3xdtN\sqrt{h}[K_{ij}K^{ij}-\lambda K^2-2\Lambda+R-g_2M_{pl}^{-2}R^2-g_3M_{pl}^{-2}R_{ij}R^{ij}\\
-g_4M_{pl}^{-4}R^3-g_5M_{pl}^{-4}RR^i_jR^j_i-g_6M_{pl}^{-4}R^i_jR^j_kR^k_i-g_7M_{pl}^{-4}R\nabla^2R-g_8M_{pl}^{-4}\nabla_iR_{jk}\nabla^iR^{jk}],
\end{aligned}
\end{equation}
with $N$ the shift function, $h$ the determinant of the induced metric on spatial hypersurface, $K_{ij}$ the extrinsic curvature, and $\l$, $g_2$, ... $g_{8}$ parameters of the theory. Quantum cosmology based on this theory of gravity has been given in \cite{Bertolami:2011ka}. The corresponding WD equation is
\begin{equation}
\la{lalala}
\left(\frac{\partial^2}{\partial a^2}+\frac{q}{a}\frac{\partial}{\partial a}-g_Ca^2+g_{\Lambda}a^4+g_r+\frac{g_s}{a^2}\right)\Psi(a)=0,
\end{equation}
with $g_C$, $g_{\Lambda}$, $g_r$ and $g_s$ defined as
\begin{equation}
\begin{aligned}
g_C=\frac{2}{3\lambda-1},&\qquad g_{\Lambda}=\frac{\Lambda M_{pl}^{-2}}{18\pi^2(3\lambda-1)^2},\\
g_r=24\pi^2(3g_2+g_3),& \qquad g_s=288\pi^4(3\lambda-1)(9g_4+3g_5+g_6).
\end{aligned}
\end{equation}
Recalling discussions shown in Sec.(\ref{section2}), it is easy to figure out that, if the normal ordering ambiguity factor $q$ is chosen appropriately in the domain of $(-1,3)$, then a nonzero cosmological constant ($g_\L\neq0$) is required to ensure the normalizability of the cosmological wave function.

\section{Conclusion}

Investigating dark energy from the perspective of quantum cosmology, it had been found that only dark energy can lead to a normalizable cosmological wave function \cite{Huang:2015eua}. This conclusion is based on the assumption that dark energy is an exotic energy component. Nevertheless, at the classical level, the accelerating expansion of the universe can also be attributed to an appropriate modification of general relativity at cosmological scale. The present work aims to figure out that whether the conclusion found in \cite{Huang:2015eua} can also stand well in modified gravities.

General speaking, WD equation derived from modified gravities can differ markedly from the ordinary one (derived from GR) which may make it is difficult to answer the above question. For dRGT gravity, it is shown that the dynamical part of the corresponding WD equation changes its form dramatically (see (\ref{equ:WDW_o_MG})). Furthermore, WD equations derived form EiBI and Ho$\check{\text{r}}$ava-Lifshitz gravity introduce new potential terms (see (\ref{equ:WDW_f_EiBI}) and (\ref{lalala})). Despite of these nontrivial variations in the form of WD equation, it is shown that the conclusion---only dark energy can lead to a normalizable cosmological wave function---can stand well in these representative theories of gravity.

The study carried out in the present work, as well as that in \cite{Huang:2015eua}, implies that there may be a tight causal relationship between dark energy and normalizability of cosmological wave function. It also indicates that investigating dark energy from the perspective of quantum cosmology may be an valuable attempt. For further study, one can notice the fact that an important kind of approach to dark energy problem is reforming general relativity at its most fundamental
level to accommodate intrinsically an accelerating expansion of the universe, such as de Sitter special relativity \cite{Aldrovandi:2006vr} and uniformly expanding vacuum conjecture \cite{Huang:2016fws}. Whether the conclusion presented here can find its counterparts in these new insights to dark energy problem needs further attention. Most importantly, this concise result---only dark energy can lead to a normalizable cosmological wave function--- definitely demands an explanation in our opinion. This will be an interesting direction for further work.

%\Acknowledgments
%\noindent
%{\large{\bf Acknowledgments}}

%\vspace*{5mm}

\end{document}